\documentclass[sigconf,natbib=true]{acmart}

\AtBeginDocument{%
  }

\setcopyright{acmlicensed}
\copyrightyear{2024}
\acmYear{2024}
\acmDOI{XXXXXXX.XXXXXXX}

\acmConference[(tech report 2024)]{Make sure to enter the correct
  conference title from your rights confirmation emai}{}{}
\acmISBN{978-1-4503-XXXX-X/18/06}

\newcommand{\token}[1]{\texttt{#1}}
\newcommand{\CLS}{\token{{[CLS]}}}
\newcommand{\SEP}{\token{{[SEP]}}}
\newcommand{\MASK}{\token{[MASK]}}
\newcommand{\Q}{\token{{[Q]}}}

\usepackage{subfig}

\begin{document}

\title{ColBERT {\MASK} Tokens Perform Term Weighting and Exhibit Cyclic Contextualization }
\title{ColBERT {\MASK} Tokens Weight Non-{\MASK} Tokens and Have a Cyclic Contextualization Pattern Based on Position }
\title{ColBERT {\MASK} Tokens Weight Non-{\MASK} Terms and Repeat Across Input Position Past the Input Length for Scoring}
\title{ColBERT's {\MASK}-based Query Augmentation:\\Effects of Quadrupling the Query Input Length}
\author{Ben Giacalone}
\email{bsg8294@rit.edu}
\affiliation{%
  \institution{Rochester Institute of Technology}
  \streetaddress{1 Lomb Memorial Dr}
  \city{Rochester}
  \state{New York}
  \country{USA}
  \postcode{14623}
}

\author{Richard Zanibbi}
\email{rxzvcs@rit.edu}
\affiliation{%
  \institution{Rochester Institute of Technology}
  \streetaddress{1 Lomb Memorial Dr}
  \city{Rochester}
  \state{New York}
  \country{USA}
  \postcode{14623}
}

\begin{abstract}
A unique aspect of ColBERT is its use of 
{\MASK} tokens in queries to score documents (\emph{query augmentation)}.
Prior work shows {\MASK} tokens weighting non-{\MASK} query terms, emphasizing certain tokens over others
, rather than introducing whole new terms as initially proposed.
%
%
%
We begin by 
demonstrating that a term weighting behavior 
previously reported for {\MASK} tokens 
in ColBERTv1 
holds for 
ColBERTv2.
We then examine the effect of changing the number of {\MASK} tokens from zero to up to four times past the query input length used in training, both  
for first stage retrieval, and for scoring candidates,
observing an initial decrease in performance with few {\MASK}s, a large increase when enough {\MASK}s are added to pad queries to   an average length of 32, then a plateau in performance afterwards.
Additionally, we compare baseline performance to performance when the query length is extended to 128 tokens, and find that differences are small (e.g., within 1\% on various metrics) and generally statistically insignificant, indicating performance does not collapse if ColBERT is presented with more {\MASK} tokens than expected.


\end{abstract}

\keywords{ColBERT, BERT, mask tokens, term weighting, query augmentation}


\maketitle

\section{Introduction}

\begin{figure}
    \centering
    \includegraphics[width=1.0\linewidth]{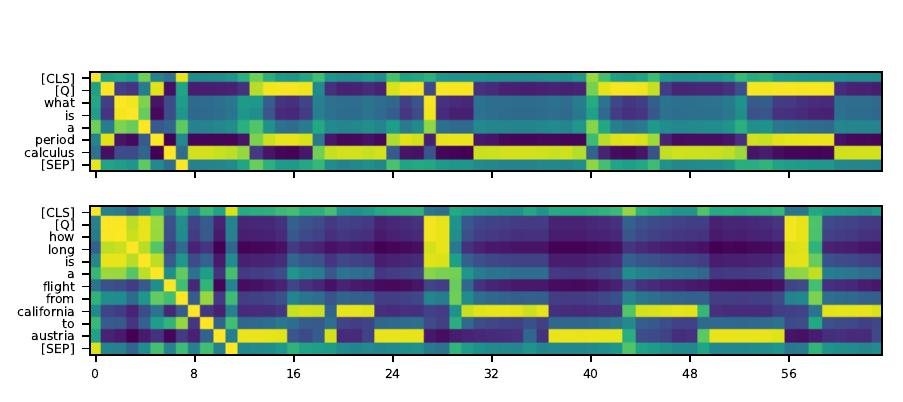}
    \caption{
    Cosine similarity of embedded tokens to each non-{\MASK} token for positions 0 through 64. 
    A cyclical pattern attending to the most relevant terms in the query (e.g. ``period'', ``calculus'', ``california'', ``austria'') can be seen, both before and after 32 tokens (the length trained with).} 
    \label{fig:weighting_pattern}
\end{figure}

ColBERT~\cite{ColBERT}'s use of multiple token embedding vectors supports fine-grained matching between queries and documents. 
The model ranks documents by 
adding the maximum similarity of a document token embedding
to each query token embedding, as shown in Equation~\ref{eq:maxsim}. This greedy alignment of query to document token embeddings 
has been dubbed \emph{MaxSim}.
\begin{equation}
\label{eq:maxsim}
S_{d,q} := \sum_{i \in [|E_q|]} \max_{j \in [|E_d|]} E_{q_i} \cdot E_{d_j}^T
\end{equation}
Here the score for document $d$ given query $q$, is computed from the set of query and document token embeddings ($E_q$ and $E_d$, respectively). Embeddings are produced by a BERT-based model~\cite{BERT} finetuned with ColBERT's training objective. For queries, 
ColBERT 
prepends a {\Q} token to indicate a query is being contextualized, and surrounds the tokens with {\CLS} and {\SEP} tokens
to indicate the beginning and ending of a passage. 
Finally, the query is padded with {\MASK} tokens up to a maximum length of 32 tokens.
Augmenting the query with
{\MASK} rather than 
standard {\texttt{[PAD]}} tokens is key to ColBERT's effectiveness. 

In Khattab and Zaharia's original ColBERT paper \cite{ColBERT}, they show using augmentation with {\MASK} tokens increases 
MRR@10 on MS MARCO~\cite{bajaj2016ms}.
Their rationale is that {\MASK} tokens help
introduce new terms to the query, and 
reweight other query terms.
However, later work suggests that {\MASK} tokens primarily weight other tokens in the query, as summarized in Section \ref{sec:related}. In this paper we present new experiments to obtain additional insight into how query augmentation maps {\MASK}s into the contextualized token embedding space. 
We consider two main research questions:

\begin{description}
\item[\small RQ1.] Do {\MASK} tokens primarily weight non-{\MASK} tokens in a query when using ColBERTv2?
\item[\small RQ2.] Does effectiveness increase with the number of {\MASK}s, up to four times the number ColBERT has been trained with?
\end{description}


\section{Related Work}
\label{sec:related}


Prior work has analyzed how ColBERT contextualizes tokens.
\citet{formal2021whitebox} focused their analysis on query text tokens, using both a model trained with {\MASK}s and a model finetuned without {\MASK}s during ranking. They found that query text tokens implicitly capture term importance, because terms with higher IDF tend to produce more exact matches, and change their embedded representation less. When using a model that was finetuned to not use {\MASK}s, this effect was even more apparent.

\citet{wang2023prf} considered 
whether 
{\MASK}s in ColBERT actually add new terms to the query, as \citet{ColBERT} proposed in their original paper. They found that it did not, and presented an IDF-based approach for adding new terms to the query.
%
In the same paper, the authors show that {\MASK} tokens tend to cluster around items already present the query, rather than produce novel query terms, necessitating an approach such as pseudorelevance feedback to add additional query terms. 

More recently, \citet{giacalone2024mask} remapped contextualized {\MASK} embeddings to their nearest non-{\MASK} embedding (i.e. {\CLS}, {\SEP}, {\Q}, and the query text tokens), and found no significant difference in MRR@10, nDCG@10/@1000. However, a significant \emph{increase} in MAP was observed both when remapping {\MASK} vectors to their nearest query text token vector, and when remapping {\MASK} vectors to their nearest non-{\MASK} token vector.
While interesting, a shortcoming is that their expeirments consider only ColBERTv1, instead of the more effective ColBERTv2~\cite{ColBERTv2}. 

ColBERTv2 uses a more powerful cross-encoding ranker to generate positives and negatives to train with, while ColBERTv1 uses labelled positives and random negatives. This results in an almost 4\% gain in MRR@10 on the MS MARCO dev set, allowing it to compete with newer dense retriever models that take advantage of distillation (e.g. PAIR~\cite{ren2021pair}).
This may change the behavior of how {\MASK}s interact with non-{\MASK} tokens. In our first experiment, we attempt to replicate \citet{giacalone2024mask}'s results using ColBERTv2.

\citet{Tonellotto2021querypruning} demonstrated that the number of query token embeddings required for \emph{initial retrieval} in ColBERT can be reduced to as little as 3 by pruning terms frequently present in the collection. They found that {\MASK}s tend to add less documents to the initial set of documents retrieved, since {\MASK}s tend to be very similar to existing terms in the documents.
Similar to this paper, in our second set of experiments we perturb the model by modifying the number of {\MASK} tokens available.

\section{Methodology}

We run our experiments using PyTerrier~\cite{macdonald2021pyterrier}, which contains advanced bindings for ColBERT. Into this framework we load the ColBERT v2~\cite{ColBERTv2} checkpoint provided by the ColBERT team.\footnote{\url{https://downloads.cs.stanford.edu/nlp/data/colbert/colbertv2/colbertv2.0.tar.gz}} We confirmed that this checkpoint was trained using the default query length of 32, and that {\MASK}s had their attention scores zeroed out during training (i.e. no token can attend to a {\MASK} token during self attention).
PyTerrier officially supports only ColBERTv1, but we have verified that the keys PyTrrier expects are also present in the our v2 checkpoint. 

We do not use v2's index compression, but we believe this is acceptable, since this is not 
a core feature of the retrieval model. 
Using the uncompressed index does
slightly change performance on MS MARCO from the official metrics. On the MS MARCO dev set, we obtained an MRR@10 of 39.8, Recall@50 of 86.0, and Recall@1000 of 96.2, compared to the official reported metrics of MRR@10 of 39.7, R@50 of 86.8, and R@1000 of 98.4. We suspect this increase in Recall is due to some terms becoming more similar when index compression is applied.

We run our experiments on a server with 4 Intel Xeon E5-2667v4 CPUs, 4 NVIDIA RTX2080-Ti GPUs, and 512~GB RAM. 
We use two datasets from \citet{giacalone2024mask}:
\begin{enumerate}
    \item MS MARCO~\cite{bajaj2016ms}'s passage retrieval dev set (8.8 million documents, 1 million queries, binary relevance judgements). Each query has at most 1 matching document.
    \item A dataset combining queries from the TREC 2019~\cite{trec2019} and 2020~\cite{trec2020} deep passage retrieval task (99 queries, graded relevance judgements). Collection is the same as MS MARCO.
\end{enumerate}
As in \cite{giacalone2024mask} we use MS MARCO when relevance grades are unimportant important, and use the latter when it is, and consider different relevance levels during evaluation.
Additionally, for RQ2, we also use the TREC COVID dataset~\cite{trecCOVID} in addition to the TREC 2019-2020 dataset. This dataset contains 50 queries with graded relevance judgements from 0 to 3. Note that we use the CORD-19 variant~\cite{cord19} instead of the BEIR variant~\cite{beir} used in the ColBERTv2 paper; thus our baseline measurement differs from the officially reported figure.


\paragraph{\bf RQ1: Do [MASK] tokens primarily weight non-[MASK] tokens in a query when using ColBERTv2?}
We reproduce the experiments from \citet{giacalone2024mask} on ColBERT v2,  using the TREC 2019-2020 collection.
In the first experiment, we compare a baseline of the standard retrieval pipeline against three conditions where certain token embeddings are replaced with others:
1. We remap \emph{all} structural token embeddings (i.e. {\CLS}, {\SEP}, {\Q}, {\MASK}) to their nearest query text token embedding.
2. We remap {\MASK} tokens to their nearest non-MASK token (i.e. {\CLS}, {\SEP}, {\Q}, query text tokens).
3. We remap {\MASK} tokens to their nearest query text embedding, but leave other structural token embeddings (i.e. {\CLS}, {\SEP}, {\Q}) alone.

In the second experiment, we modify all queries in the TREC 2019-2020 collection with a length of 3-8 tokens that start with ``what is'' by moving these two tokens to the end of the query and swapping their positions (e.g. ``\emph{what is} love'' becomes ``love \emph{is what}'').
As indicated in the original paper, this 
avoids changing query semantics, 
while shifting the position of every query token.
We check the change in cosine distance for {\CLS}, {\SEP}, {\Q}, the first and third query text token, and the 13th and 32nd token in the query, which are guaranteed to be {\MASK} tokens.
As a baseline, we repeat the same experiment without requiring queries to start with "what is", 
possibly generating nonsense (e.g. ``\emph{cost of} swim spa'' becomes ``swim spa \emph{of cost}'').

\paragraph{\bf RQ2: Does effectiveness increase with the number of {\MASK}s, up to four times the number ColBERT has been trained with?}
As shown in Figure~\ref{fig:weighting_pattern}, when extending the maximum length of a query past the 32 token window it was trained with, we see a repeating pattern of cosine similarities between {\MASK} and non-{\MASK} tokens.
It appears that BERT keeps outputting the same weighting pattern for longer query lengths.
A natural question then, is how ColBERT fares when the maximum query length is increased, and {\MASK}-based term weighting dominates document scoring.
%
One may be wary of the unintentional effects of changing {\MASK} counts this way. For instance, could adding an extra {\MASK} to the end of a query cause the previous {\MASK}s, or even the query text tokens, to change their representations in response?

An easily missed detail about ColBERT is that it treats {\MASK} and non-{\MASK} tokens differently during the contextualization process ---
{\MASK} tokens cannot be attended to during self attention\footnote{To our knowledge, this has not 
been reported 
in the 
ColBERT papers.}.
This has two interesting consequences.
One, adding or subtracting {\MASK}s cannot affect how non-{\MASK} tokens are contextualized.
Non-{\MASK} tokens cannot attend to {\MASK} tokens, thus removing {\MASK}s from the query entirely will not change any of the non-{\MASK} representations.
Two, each {\MASK} token's computed representation cannot be affected by the existence of \emph{other} {\MASK} tokens.
Each {\MASK} token can only look at the query and itself,
thus, the only change to scoring when adding or removing a {\MASK} token is the existence of the token's score. In other words, other tokens cannot change their representations in response to to different numbers of {\MASK} tokens.

In our second experiment,
we vary the maximum length of the query from 0 to 96 in steps of two, and measure the resulting performance on TREC 2019-2020. 
Since we start from a length of 0, we hypothesize that performance will initially increase greatly with each additional {\MASK}, reflecting the importance of query augmentation.
Performance will then plateau, 
even as more {\MASK}s are added than seen during training, as the {\MASK}s repeatedly perform a similar term weighting.

Separately, we report nDCG@10 and nDCG@1000 when the maximum query length is set to 32 to 128, to identify the effect of increasing the total number of tokens seen for each query. In addition to the TREC 2019-2020 dataset, we also use the TREC COVID dataset for this experiment.

ColBERT performs ranking in two phases: an initial set retrieval phase, where documents with at least one embedding very similar to a query embedding are fetched, and a subsequent reranking phase, where documents are reranked by MaxSim. In all experiments, we report metrics for (1) only initial set retrieval is modified, (2) only reranking is modified, and (3) both phases are modified. 


\section{Results}

\begin{table}[!t]
    \centering
    \caption{Replacing structural token embeddings by other query token embeddings (TREC 2019-2020, RQ1).
    Maximum values are in bold; significant differences from ``None'' are shown with a dagger ($p<0.05$, Bonferroni-corrected $t$-tests).}

    \resizebox{0.85\linewidth}{!}{
    \begin{tabular}{l | p{0.5in} p{0.5in} p{0.5in} p{0.5in} }
      \toprule
    & \multicolumn{4}{c}{\sc *ColBERTv2: ~Structural Token Remapping}\\
         \multicolumn{1}{l|}{{\sc Metric}} & {\bf None} & {\bf All {\tt [X]} $\rightarrow$ Text } & {\bf {\MASK} $\rightarrow$ Text } & {\bf {\MASK} $\rightarrow$ Str. \& Text}  \\
         \hline
         \textbf{Binary Rel.}\\
         MAP(rel$\geq$1) & \textbf{0.514} & $\dagger$0.496 & $\dagger$0.508 & 0.510 \\
         MRR(rel$\geq$1)@10 & \textbf{0.964} & 0.958 & 0.959 & 0.960 \\
         \hline
         MAP(rel$\geq$2) & \textbf{0.502} & $\dagger$0.489 & 0.496 & 0.498 \\
         MRR(rel$\geq$2)@10 & 0.870 & 0.871 & \textbf{0.888} & 0.874 \\
         \hline
         MAP(rel$\geq$3) & \textbf{0.395} & 0.388 & 0.387 & 0.391 \\
         MRR(rel$\geq$3)@10 & \textbf{0.616} & 0.593 & 0.598 & 0.605 \\
         \midrule
         \textbf{Graded Rel.}\\
         nDCG@10 & \textbf{0.749} & $\dagger$0.733 & 0.741 & 0.745 \\
         nDCG@1000 & \textbf{0.712} & $\dagger$0.691 & $\dagger$\,0.702 & $\dagger$0.703 \\
        \bottomrule
    \end{tabular}
    }
    \label{tab:ecir_mask}
\end{table}

\paragraph{\bf RQ1: Do [MASK] tokens primarily weight non-[MASK] tokens in a query when using ColBERTv2}
For the {\MASK} remapping experiment, we see that on ColBERTv2, remapping {\MASK}s causes a consistent decrease in performance (see Table \ref{tab:ecir_mask}). For nDCG@1000, all conditions are significantly worse than the baseline. The ``All [X] $\rightarrow$ Text'' condition performs worse than any other condition, many times being significantly worse than the baseline. The ``{\MASK} $\rightarrow$ Str. \& Text'' condition performs best of the three conditions. This is both consistent with the ColBERTv1 results from \citet{giacalone2024mask}, and provides more evidence for that {\MASK} embeddings simply select all non-{\MASK}s as candidates for term weighting.

\begin{figure}
    \centering
    {\sc ColBERTv1}
    \includegraphics[width=1.\linewidth]{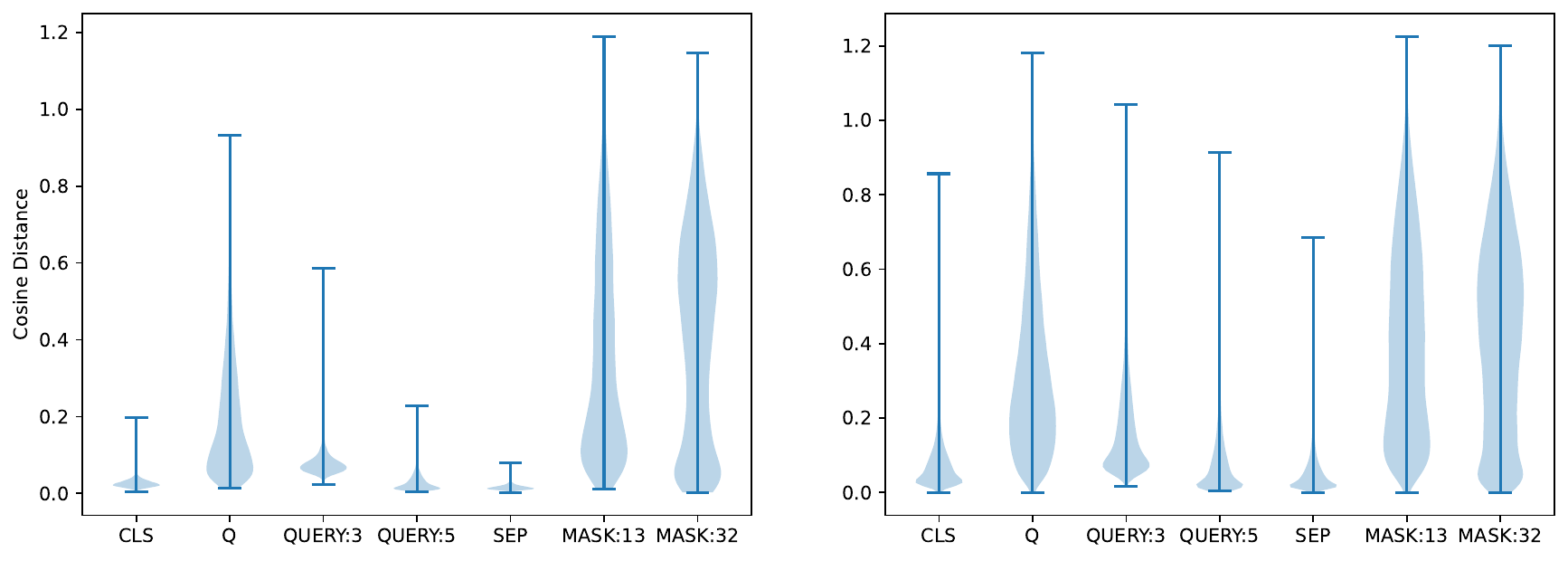}

    {\sc ColBERTv2}
    \includegraphics[width=1.\linewidth]{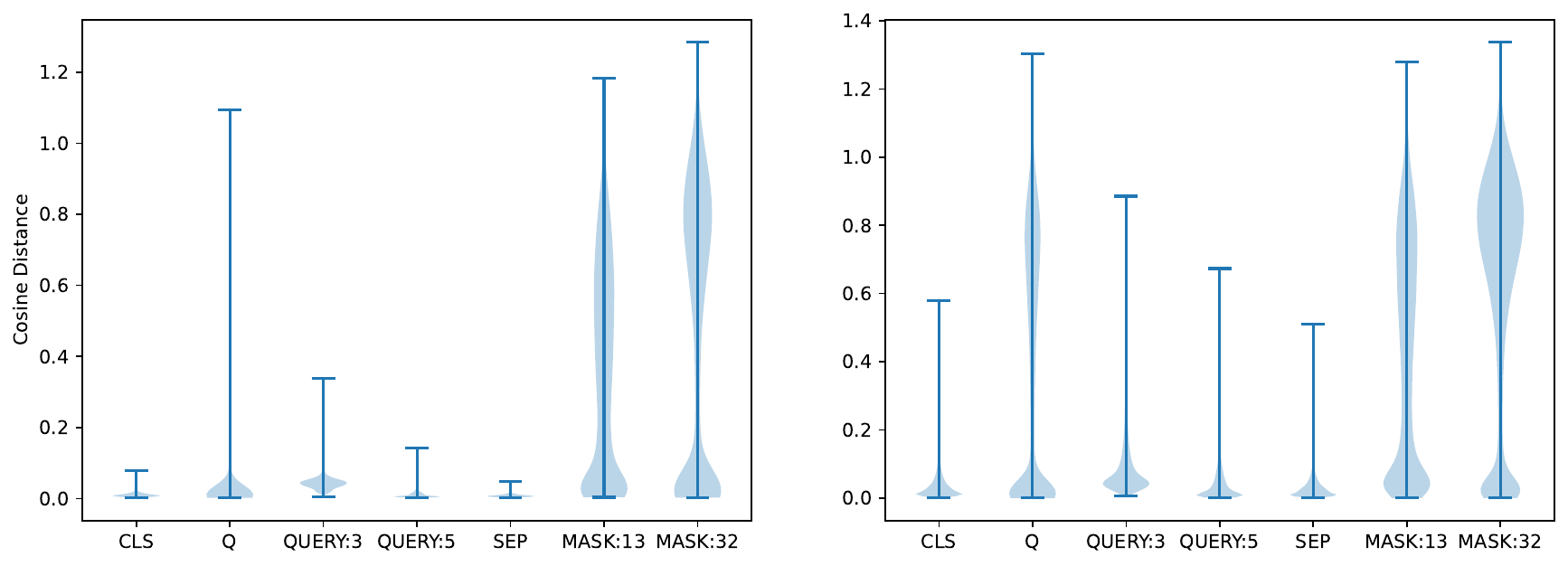}
    \caption{Left column: Cosine distance after tokens are switched from "what is" to "is what"  in ColBERTv1 vs. ColBERTv2. We see the same trend of {\Q} and {\MASK} tokens having the most shifting, and an overall increase in shifting when ``what is'' is not a requirement (right column). The contrast between {\Q} and {\MASK} versus other tokens is more apparent in ColBERTv2 than ColBERTv1.}
    \label{fig:cosine_dist_v2}
\end{figure}

For the query shift experiment shown in Figure~\ref{fig:cosine_dist_v2}, we see the same pattern reported in \citet{giacalone2024mask}: {\Q} and {\MASK} tokens vary greatly after ``what is'' is swapped and moved, while {\CLS}, {\SEP}, and query text tokens do not change nearly as much. In fact, with ColBERTv2, this difference is even starker. Given that this is a pattern that has now manifested itself across two separately trained checkpoints, with two different training objectives, we suspect that the {\Q} token performs a similar function to {\MASK} tokens -- adding weight to certain tokens to influence scoring.

This would also explain the pattern demonstrated by the {\Q} token in Figure~\ref{fig:weighting_pattern}, where {\MASK}s that are very similar to the {\Q} token are always also very similar to some other token. When we visualized several different queries using the same visualization shown in Figure~\ref{fig:weighting_pattern}, we saw that {\Q} was the only non-{\MASK} structural token consistently very similar to query text tokens.


\begin{table}[!t]
    \centering
    \caption{Changing the maximum length of queries from 32 to 128 with {\MASK} padding.
    Maximum values are in bold; significant differences from ``32'' are shown with a dagger ($p<0.05$, Bonferroni-corrected $t$-tests).}

    \resizebox{1.0\linewidth}{!}{
    \begin{tabular}{l | p{0.5in} p{0.5in} | p{0.5in} p{0.5in} }
      \toprule
    & \multicolumn{2}{c}{\sc  TREC 2019-2020} & \multicolumn{2}{c}{\sc TREC COVID}\\
         \multicolumn{1}{l|}{{\sc Metric}} & 32 & 128 & 32 & 128 \\
         \hline
         \bf{Only Set Retrieval} & & & & \\
         nDCG@10 & 0.749 & 0.749 & 0.612 & \bf{0.616} \\
         nDCG@1000 & 0.712 & \bf{0.717} & 0.343 & \bf{$\dagger$0.350} \\
         \hline
         \bf{Only Reranking} & & & & \\
         nDCG@10 & \bf{0.749} & 0.739 & 0.612 & \bf{0.640} \\
         nDCG@1000 & \bf{0.712} & 0.707 & 0.343 & \bf{0.349} \\
         \hline
         \bf{Set Retrieval and Reranking} & & & & \\
         nDCG@10 & \bf{0.749} & 0.743 & 0.612 & \bf{0.643} \\
         nDCG@1000 & 0.712 & 0.712 & 0.343 & \bf{0.355} \\
        \bottomrule
    \end{tabular}
    }
    \label{tab:vary_query_maxlen}
\end{table}

\paragraph{\bf RQ2: Does effectiveness increase as the number of {\MASK}s increases up to four times the number ColBERT has been trained with?}
In Table~\ref{tab:vary_query_maxlen}, we see nDCG@/@1000 on both TREC 2019-2020 and TREC COVID as we vary the maximum query length.
We first focus on the results from the TREC 2019-2020 dataset. Modifying only set retrieval causes a minor increase in nDCG@1000, but appears to have no effect on nDCG@10, likely due to baseline set retrieval already retrieving most relevant documents.
Modifying only reranking on TREC 2019-2020 causes both nDCG@10/@1000 to decrease. When modifying both phases, nDCG@10 very slightly increases, but nDCG@1000 does not change, likely due to the increase from set retrieval and the decrease from reranking negating each other.
Ultimately, all changes observed on TREC 2019-2020 are small, and we never saw an increase or decrease greater than 1\%, nor did we observe any statistically significant $p$-values when performing Bonferroni-corrected $t$-tests.

On the TREC COVID dataset, we see an increase in nDCG@/@1000 as we increase the length of the query to 128 tokens, for both reranking and set retrieval. These changes are still very small, in the range of 1-3\%. The increase in nDCG@1000, however, is statistically significant.

A possible reason for this difference in behavior between TREC 2019-2020 and COVID is that the former dataset has less tokens per query on average compared to the latter (9.68 versus 13.92 tokens), potentially causing certain queries to be incompletely weighted when using only 32 tokens. 


\begin{figure}
    \centering
    \includegraphics[width=0.95\linewidth]{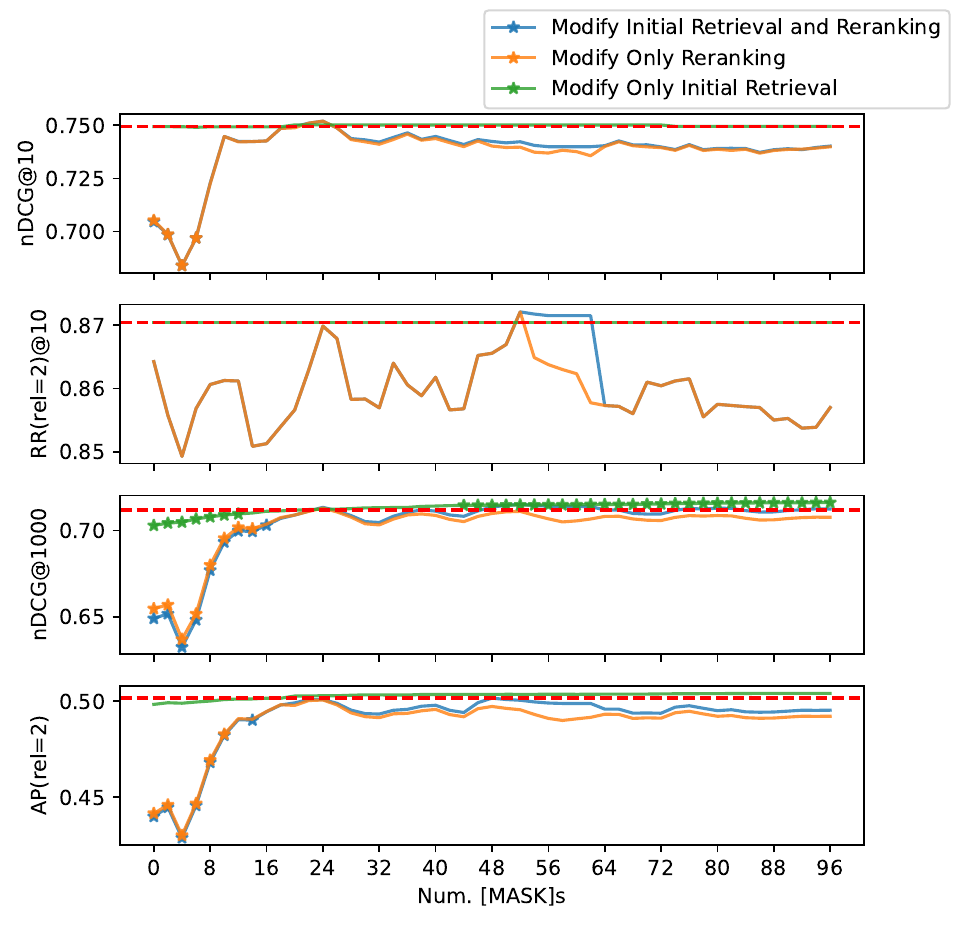}
    \vspace{-0.25in}
    \caption{nDCG@10, MRR(rel$\geq$2)@10, nDCG@1000, and MAP(rel$\geq$2) increasing number of {\MASK} tokens from 0 to 96 on TREC 2019-2020. The red line shows a standard length of 32 total tokens. Significant differences from the baseline indicated with a start (Bonferroni correction, $p<0.05$).}
    \label{fig:abs_mask_count}
\end{figure}

In Figure~\ref{fig:abs_mask_count}, we see nDCG@10/@1000, MRR(rel$\geq$2)@10, and MAP(rel$\geq$2) as we vary the number of {\MASK} tokens each query has.
For most of the metrics, moving from 0 to 4 {\MASK}s appears to actually have a detrimental affect, indicating only using a couple {\MASK} tokens is worse than none at all.
From 4 to $\sim$24 {\MASK}s, however, we see a sharp increase in nDCG@10/@1000 and MAP(rel$\geq$2).
This peak coincides with the point where on average, queries have an overall length of 32 (i.e., the input size used for training).
From there on, there is a slight decrease across all metrics, which we expect from the results of the previous experiment.
However, 
we also
can see that despite this slight reduction in performance, it is still far better than not having any {\MASK}s at all.

It appears that as more {\MASK}s are used on this collection, performance tends to converge to slightly below the baseline. As seen in Figure~\ref{fig:abs_mask_count}, using 8 {\MASK}s or less causes a statistically significant reduction in performance, while using more than that results in performance that is not significantly different from the baseline.
Also,
while increasing the number of {\MASK} tokens from 0 to 96, RR(rel$\geq$2)@10 does not change in a statistically significant way. For the TREC 2019-2020 dataset, query augmentation does not significantly impact RR(rel$\geq$2)@10.

\section{Conclusion}

The unconventional decision to have ColBERT integrate the padding token used for queries ({\MASK}) directly into its scoring mechanism has resulted in state of the art performance.
Padding with {\MASK} tokens has been demonstrated to act analogous to term weighting, making it more important for documents to match against some terms than others.
An interesting aspect of {\MASK} representations is that they form a repeating pattern, even when expanding the query past the maximum query length trained with.

We were able to 
confirm the findings of \citet{giacalone2024mask} on ColBERTv1, showing that even with ColBERTv2, remapping {\MASK}s to their nearest non-{\MASK} generally produces non-significant differences in effectiveness metrics, and that {\MASK}s are much more sensitive to token order than {\CLS}, {\SEP}, and even query text tokens.
We also found that a partial term weighting using fewer {\MASK} tokens than used in trained causes effectiveness to decrease, 
i.e. using no {\MASK}s performs better than using a small number of {\MASK}s.
Increasing the number of {\MASK}s from this low point to the amount trained with causes performance to shoot up.
Afterwards, performance slightly reduces as {\MASK}s are added across most metrics, but still performs much better than not using {\MASK}s at all.

Overall, though there is a slight drop in performance, ColBERT's {\MASK}-based term weighting strategy performs well past the maximum query length it was trained with, converging to near baseline levels as the size of the query input increases.



\bibliographystyle{ACM-Reference-Format}
\bibliography{bibliography}

\end{document}